\documentclass[preprint]{aastex631}
\shorttitle{Improve energy resolution with gain correction}
\shortauthors{Wu et al.}
\graphicspath{{./}{figure/}}
\begin{document}

%\title{The improved near-Fano-limit X-ray energy resolution of a scientific CMOS detector at room temperature}

\title{Improving the X-ray energy resolution of a scientific CMOS detector by pixel-level gain correction}

\author{Qinyu Wu}
\affiliation{National Astronomical Observatories, Chinese Academy of Sciences \\
20A Datun Road, Chaoyang District \\
Beijing 100101, China}
\affiliation{School of Astronomy and Space Science, University of Chinese Academy of Sciences \\
19A Yuquan Road, Shĳingshan District \\
Beijing 100049, China}

\author{Zhixing Ling}
\affiliation{National Astronomical Observatories, Chinese Academy of Sciences \\
20A Datun Road, Chaoyang District \\
Beijing 100101, China}
\affiliation{School of Astronomy and Space Science, University of Chinese Academy of Sciences \\
19A Yuquan Road, Shĳingshan District \\
Beijing 100049, China}
\correspondingauthor{Zhixing Ling}
\email{lingzhixing@nao.cas.cn}

\author{Xinyang Wang}
\affiliation{Gpixel Inc. \\
No. 588 Yingkou Road, Economical and Technological Development Zone \\
Changchun 130033, China}

\author{Chen Zhang}
\affiliation{National Astronomical Observatories, Chinese Academy of Sciences \\
20A Datun Road, Chaoyang District \\
Beijing 100101, China}
\affiliation{School of Astronomy and Space Science, University of Chinese Academy of Sciences \\
19A Yuquan Road, Shĳingshan District \\
Beijing 100049, China}

\author{Weimin Yuan}
\affiliation{National Astronomical Observatories, Chinese Academy of Sciences \\
20A Datun Road, Chaoyang District \\
Beijing 100101, China}
\affiliation{School of Astronomy and Space Science, University of Chinese Academy of Sciences \\
19A Yuquan Road, Shĳingshan District \\
Beijing 100049, China}

\author{Shuang-Nan Zhang}
\affiliation{National Astronomical Observatories, Chinese Academy of Sciences \\
20A Datun Road, Chaoyang District \\
Beijing 100101, China}
\affiliation{School of Astronomy and Space Science, University of Chinese Academy of Sciences \\
19A Yuquan Road, Shĳingshan District \\
Beijing 100049, China}
\affiliation{Institute of High Energy Physics, Chinese Academy of Sciences \\
19B Yuquan Road, Shĳingshan District \\
Beijing 100049, China}



\begin{abstract}

Scientific Complementary Metal Oxide Semiconductor (sCMOS) sensors are finding increasingly more applications in astronomical observations, thanks to their advantages over charge-coupled devices (CCDs) such as a higher readout frame rate, higher radiation tolerance, and higher working temperature. In this work, we investigate the performance at the individual pixel level of a large-format sCMOS sensor, GSENSE\-1516\-BSI, which has $4096\times4096$ pixels, each of $\rm{15\ \mu m}$ in size. To achieve this, three areas on the sCMOS sensor, each consisting of $99\times99$ pixels, are chosen for the experiment. The readout noise, conversion gain and energy resolutions of the individual pixels in these areas are measured from a large number (more than 25,000) of X-ray events accumulated for each of the pixels through long time exposures. The energy resolution of these pixels can reach 140 eV at 6.4 keV at room temperature and shows a significant positive correlation with the readout noise. The accurate gain can also be derived individually for each of the pixels from its X-ray spectrum obtained. Variations of the gain values are found at a level of 0.56\% statistically among the 30 thousand pixels in the areas studied. With the gain of each pixel determined accurately, a precise gain correction is performed pixel by pixel in these areas, in contrast to the standardized ensemble gain used in the conventional method. In this way, we could almost completely eliminate the degradation of energy resolutions caused by gain variations among pixels. As a result, the energy resolution at room temperature can be significantly improved to 124.6 eV at 4.5 keV and 140.7 eV at 6.4 keV. This pixel-by-pixel gain correction method can be applied to all kinds of CMOS sensors, and is expected to find interesting applications in X-ray spectroscopic observations in the future. 

\end{abstract}



\section{Introduction}
\label{sect:intro}
Silicon image sensors, including charge-coupled devices (CCDs) and Complementary Metal Oxide Semiconductor (CMOS) sensors, are widely used in soft X-ray imaging and spectroscopy. In the past several decades, CCD detectors have been dominant in X-ray applications. Most modern X-ray astronomy missions, such as Chandra \citep{garmire2003advanced}, XMM-Newton \citep{struder2001european}, Suzaku \citep{suzakuCCD2007} and eROSITA \citep{meidinger2021erosita}, have chosen CCD sensors as their focal plane detectors. Recently, the performance of scientific CMOS (sCMOS) detectors has been improved considerably. Compared to CCDs, sCMOS sensors have several advantages, such as high readout frame rate, high radiation tolerance, and relaxed requirement for working temperature. These make sCMOS a favorable choice in future X-ray astronomy missions such as Einstein Probe (EP) \citep{yuan2018einstein, yuan2022the} and THESEUS \citep{heymes2020development}.

Energy resolution is an important parameter of X-ray spectrometers, and is commonly characterized by the Full Width Half Maximum (FWHM) of a given peak. The focal plane detectors of some future X-ray missions, such as Lynx \citep{gaskin2019lynx}, are required to have good soft X-ray energy resolution. For silicon sensors, Fano fluctuation in the charge production process gives the theoretical limit (Fano limit) on the energy resolution \citep{fano1947}, which is around 124 eV at 6.4 keV at room temperature. The energy resolution of Silicon Drift Detectors (SDDs) can reach 122 eV at 5.9 keV at $-60 ^\circ \rm{C}$\footnote{\url{https://www.amptek.com/products/x-ray-detectors/fastsdd-x-ray-detectors-for-xrf-eds/fastsdd-silicon-drift-detector}}. And modern CCDs can reach 130 eV at 6.4 keV under deep refrigeration of below $-70 ^\circ \rm{C}$ \citep{meidinger2021erosita}. But for CMOS sensors, the typical energy resolution is worse than that of CCDs \citep{wang2019characterization,narukage2020high,WU2022,hsiao2022xray}. Many efforts have been made to improve the energy resolution of CMOS sensors. \citet{hull2019hybrid} reported that 148 eV at 5.9 keV can be reached for a hybrid CMOS sensor, by applying pixel-by-pixel gain correction. At around $-30 ^\circ \rm{C}$, \citet{heymes2022characterisation} reached a resolution of 142 eV at 5.9 keV with a CIS221-X sensor, and a backside illuminated version of this sensor reached 153 eV at 5.9 keV at $-40 ^\circ \rm{C}$ \citep{stefanov2022a}.  

Our lab has been studying the X-ray performance of sCMOS sensors since 2015. We have shown that sCMOS sensors are applicable to X-ray astronomical observations \citep{wang2019characterization, ling2021correlogram}. Cooperating with Gpixel Inc., we proposed and fabricated an X-ray sCMOS of a large format and pixel size, and a thickened epitaxial layer. This sensor, namely GSENSE\-1516\-BSI, has a physical size of 6 cm $\times$ 6 cm, with an array of $4096\times4096$ pixels and a pixel size of $\rm{15\ \mu m}$. The epitaxial layer is $\rm{10\ \mu m}$ thick, and it is fully depleted. The frame rate is 20 fps in the current design. At $\rm{-30 ^{\circ}\!C}$, the dark current is lower than 0.02 $\rm{e^-}$/pixel/s and the readout noise is lower than 5 $\rm{e^-}$. The energy resolution reaches 180.1 eV at 5.90 keV for X-ray events with the single-pixel pattern \citep{WU2022}. 

Although the readout noise of the CMOS sensor is low, usually lower than 5 $\rm{e^-}$, the energy resolution is significantly worse than the theoretical FWHM with the readout noise taken into account. One possible explanation is that each pixel of a CMOS sensor has its own amplifier, and the conversion gain can differ from one pixel to another, resulting in degraded energy resolution. If this is the case, it is expected that careful gain correction at the individual pixel level may help improve the energy resolution. In this paper, we investigate this problem by studying the sensor properties at the pixel level, for which experiments of long time exposures are required. The experimental setup is described in section~\ref{sect:exp_setup}; the data reduction method is introduced in section~\ref{sect:data_redct}; the results about the noise, gain variation and the corrected energy resolution are shown and discussed in section~\ref{sect:result_disc}; and conclusions are summarized in section~\ref{sect:concls}.

\section{Experimental Setup}
\label{sect:exp_setup}
\begin{figure}
\begin{center}
\begin{tabular}{c}
\includegraphics[width=0.5\textwidth]{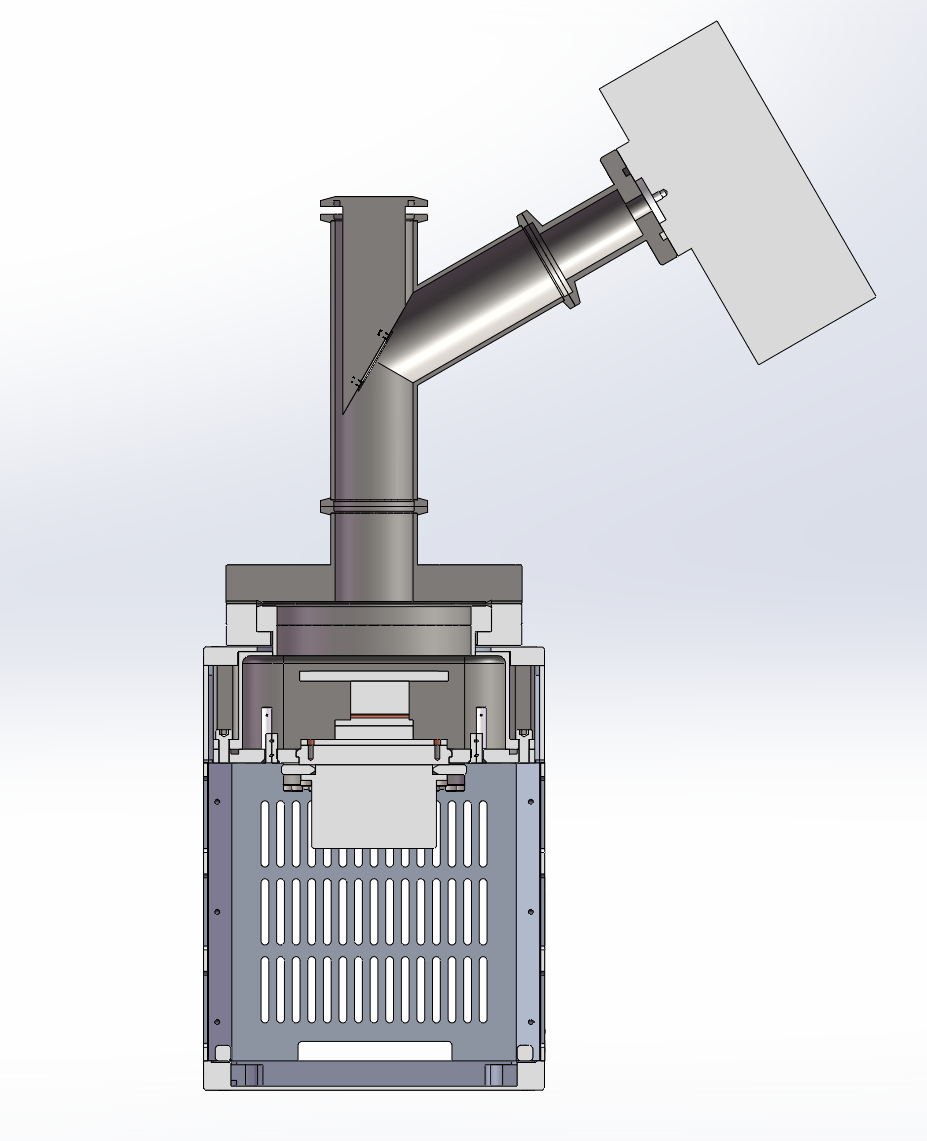}
\end{tabular}
\end{center}
\caption 
{ \label{fig:exp_device}
The experimental layout. Illuminated by initial X-rays produced by the X-ray tube at right side, the $\rm{Ti}$ target at center can produce characteristic photons. These secondary X-rays are then received and recorded by the CMOS sensor in the camera at the bottom.} 
\end{figure}

The experimental layout of our X-ray tests is shown in Fig.~\ref{fig:exp_device}. A commercial X-ray tube is placed at the right side of the device. It has a titanium anode and is operated at 25 keV. The tube can generate initial X-rays to irradiate and excite the target at the center. X-ray photons emitted from the target material are then received by the camera at the bottom. In this test, a GSENSE\-1516\-BSI sCMOS sensor is used. The camera, developed by our laboratory, contains the sensor, readout electronics and temperature control structures \citep{wang2022design}. A unique feature of the camera is that it can process images in real time and record only the extracted events, which largely reduces the storage space when the sensor works at a high frame rate. In our study, a Ti target is used. Characteristic lines of not only $\rm{Ti}$, but also $\rm{Cr}$ and $\rm{Fe}$ can be found in the spectrum, which come from the stainless steel of the surrounding structures and the walls of the device. The dark current of this chip is less than 8 $\rm{e^-}$/pixel/s or 0.4 $\rm{e^-}$/pixel/frame at room temperature, which is sufficiently low and has negligible effect on the experiment. Therefore, no cooling is needed, and all experiments are carried out at room temperature.

\begin{figure}
\begin{center}
\begin{tabular}{c}
\includegraphics[width=0.75\textwidth]{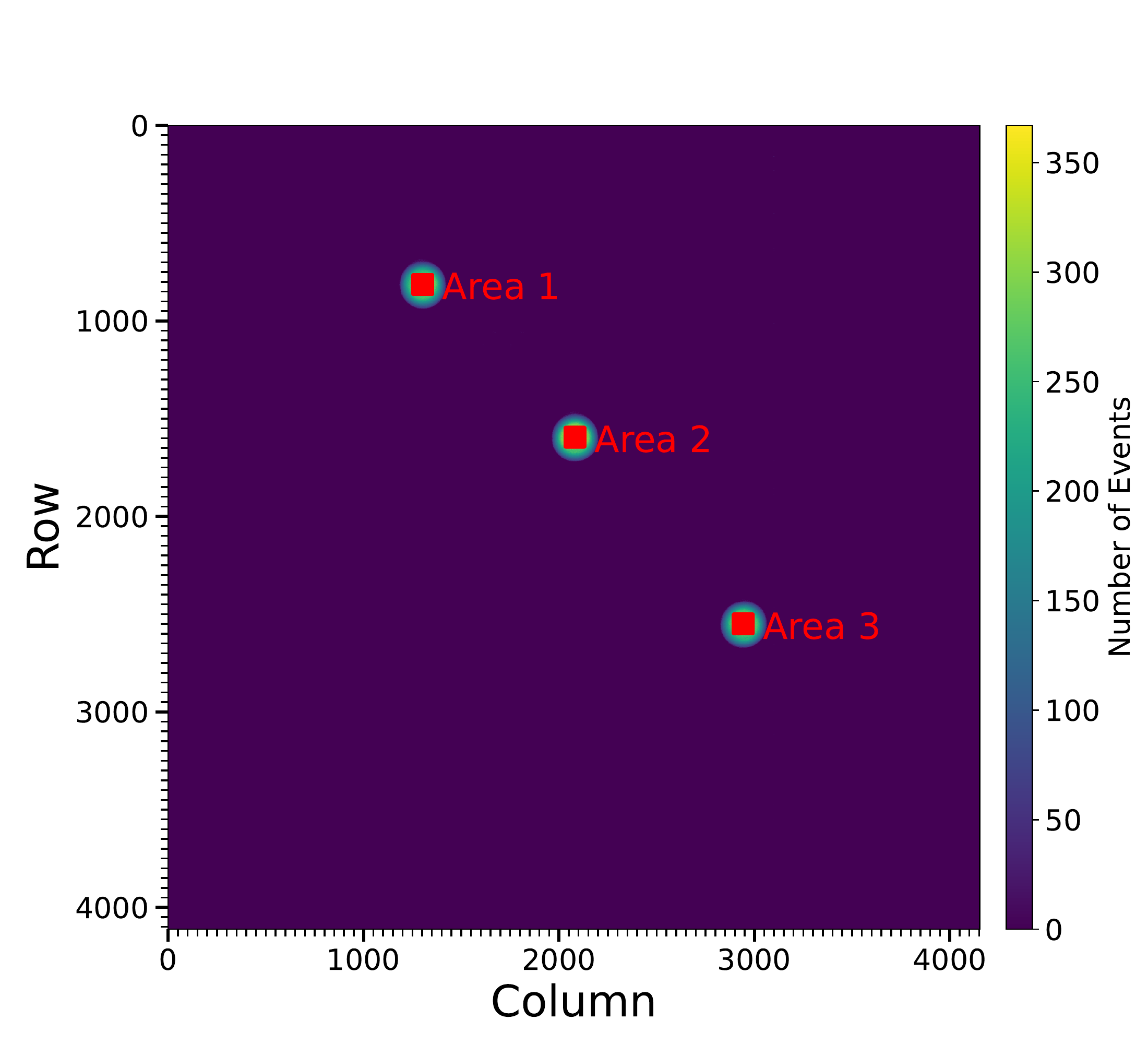}
\end{tabular}
\end{center}
\caption 
{ \label{fig:expmap}
An example exposure map. Only the pixels in three holes are exposed to X-rays. Three $\rm{99\times99}$-pixel areas (red blocks) in these holes, namely Area 1, 2 and 3, are used in the following study.} 
\end{figure}

Our goal is to accumulate at least 10,000 X-ray events on each pixel. In order to reduce the amount of experimental data, we cover the CMOS sensor with a plate, which has three 2 mm diameter holes on it. Therefore, only pixels in these three holes are exposed to X-rays. These holes are randomly located across the chip, as shown in the exposure map Fig.~\ref{fig:expmap}. So we can evaluate the non-uniformity in one area and also among these three areas, which can reflect the variations on small and large scales, respectively. Three $\rm{99\times99}$-pixel areas (red blocks on the map) in these holes, namely Area 1, 2 and 3, are chosen for the following data reduction.

\section{Data Reduction}
\label{sect:data_redct}
The sensor is operated at a frame rate of 20 Hz and the exposure time accumulated to around 220 hours in total. For each of the frames taken in the exposures, X-ray events are searched over the X-ray image and recorded in real time by the camera. The procedure was described in detail in \citet{WU2022}. If a pixel as the local maximum of its adjacent $3\times3$ pixels is above a preset threshold, ${T_{\rm event}=50\ \rm{DN}}$, this region is taken as an event. A 1-pixel event is defined as that no other pixels than the center one have a value above the split threshold, ${T_{\rm split}=15\ \rm{DN}}$, which is about 10 times the readout noise. Only 1-pixel events are selected. For each of the pixels within the 3 hole areas ($3\times99\times99=29403$ pixels in total), a number of 60,000 photons or so are accumulated. Among them, more than 40\% are 1-pixel events. This fraction can vary with the energy of incident photons, as shown in Figure 16 of \citet{WU2022}.

\begin{figure}
\begin{center}
\begin{tabular}{c}
\includegraphics[width=0.75\textwidth]{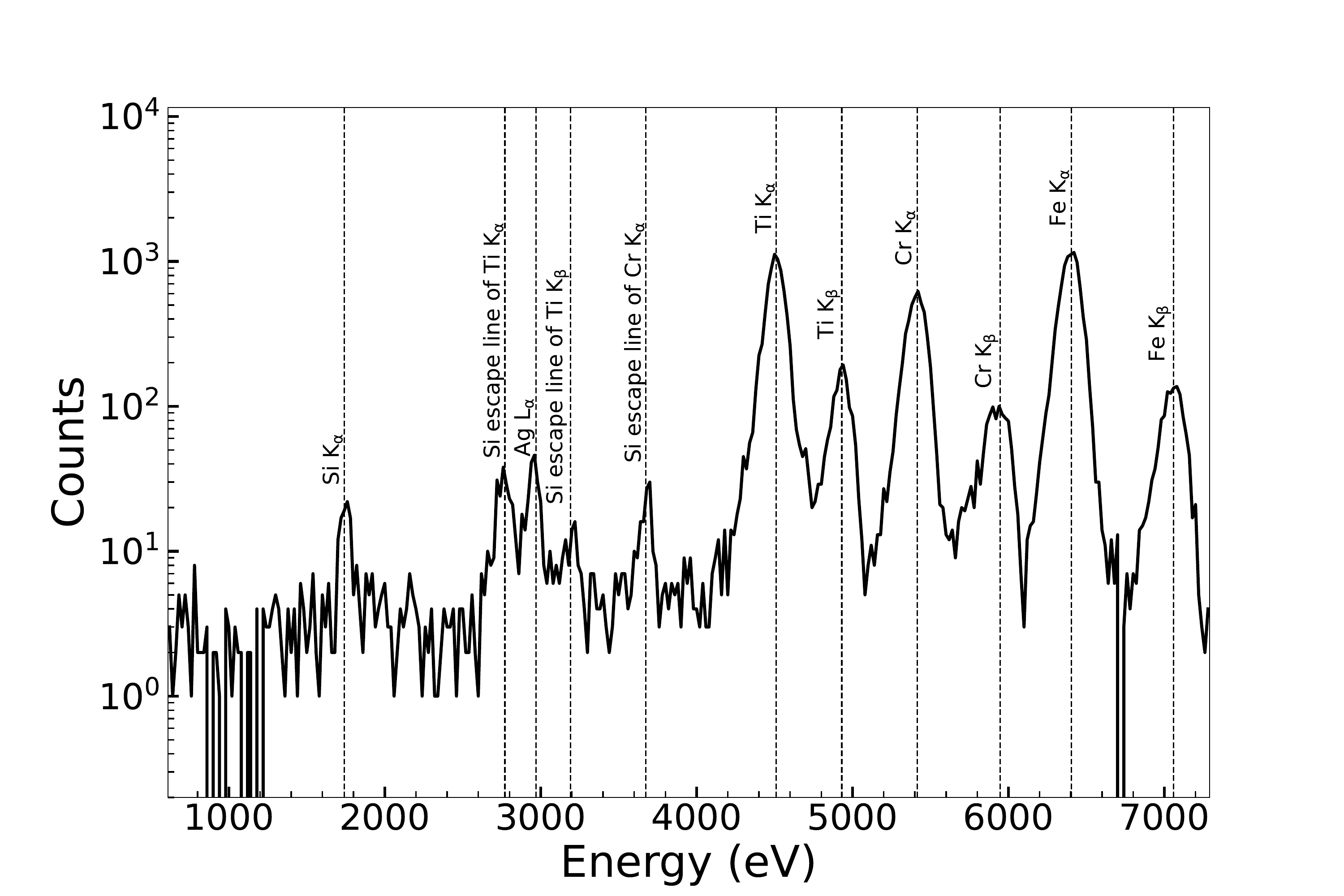}
\end{tabular}
\end{center}
\caption 
{ \label{fig:one_spec}
An example of the G0center spectrum of a single pixel. Several lines can be seen: Si $\rm{K_\alpha}$ (1.74 keV), Si escape line of Ti $\rm{K_\alpha}$ (2.77 keV), Ag $\rm{L_\alpha}$ (2.98 keV), Si escape line of Ti $\rm{K_\beta}$ (3.19 keV), Si escape line of Cr $\rm{K_\alpha}$ (3.67 keV), Ti $\rm{K_\alpha}$ (4.51 keV), Ti $\rm{K_\beta}$ (4.93 keV), Cr $\rm{K_\alpha}$ (5.41 keV), Cr $\rm{K_\beta}$ (5.95 keV), Fe $\rm{K_\alpha}$ (6.40 keV), and Fe $\rm{K_\beta}$ (7.06 keV).} 
\end{figure}

For each of the pixels, we construct an energy spectrum with the DN of the center pixel of its 1-pixel events. This spectrum is called a G0center spectrum as in \citet{ling2021correlogram} and \citet{WU2022}. Fig.~\ref{fig:one_spec} gives an example of the G0center spectrum of a single pixel. Several lines can be seen: Si $\rm{K_\alpha}$ (1.74 keV), Si escape line of Ti $\rm{K_\alpha}$ (2.77 keV), Ag $\rm{L_\alpha}$ (2.98 keV), Si escape line of Ti $\rm{K_\beta}$ (3.19 keV), Si escape line of Cr $\rm{K_\alpha}$ (3.67 keV), Ti $\rm{K_\alpha}$ (4.51 keV), Ti $\rm{K_\beta}$ (4.93 keV), Cr $\rm{K_\alpha}$ (5.41 keV), Cr $\rm{K_\beta}$ (5.95 keV), Fe $\rm{K_\alpha}$ (6.40 keV), and Fe $\rm{K_\beta}$ (7.06 keV). The Ti lines are mainly from the Ti target. The Cr and Fe lines are from the stainless steel in the structures of the experimental device. And the Ag line comes from the cover plate, which contains a small amount of silver. To determine the location and FWHM, each of the peaks is fitted with a single Gaussian function. Then we fit the relationship between the position and the known energy of the lines using the 6 strongest peaks of Ti, Cr and Fe with a linear function
\begin{equation}
\label{eq:linear}
y = a_1\times x + a_0, 
\end{equation}
where $x$ is the position of the peaks in DN and $y$ is the known energy of the emission lines in eV. The conversion gain $a_1$ in $\rm{eV/DN}$ and the intercept $a_0$ in $\rm{eV}$ of each pixel are obtained. With gain and intercept, the FWHM values in DN can be converted to energy resolutions in $\rm{eV}$ using Eq.~\ref{eq:linear}.

\section{Results and Discussion}
\label{sect:result_disc}
\subsection{Readout Noise and Dark Current}
\label{subsect:dark_prpt}
At room temperature, a series of dark experiments have been performed under different integration times, ranging from the shortest $\rm{\sim 13.92\ \mu s}$ to $100\ \rm{s}$, to measure the noise and the dark current. Thirty dark frames are recorded for each of the integration times, and the noise of each pixel is calculated as the standard deviation of these 30 values. The median value among these pixels is chosen to represent the noise of the whole frame. The readout noise $\sigma_{\rm readout}$ is 2.65 $\rm{e^{-}}$ under the shortest integration time. The noise for integration time of $50\ \rm{ms}$, $\sigma_{\rm dark50}$, increases to 2.79 $\rm{e^{-}}$ with the dark current contribution. $50\ \rm{ms}$ is the same integration time as in the X-ray exposure experiments.

For the dark current, it is calculated from a linear fit of the dark charge as a function of integration times. The overall dark current is 7.9 $\rm{e^-}$/pixel/s or 0.4 $\rm{e^-}$/pixel/frame, at room temperature. This value of dark current can just explain the difference in noise.

\subsection{Pixel-by-pixel Properties}
\label{subsect:pixel_prop}
We obtain the conversion gain and energy resolutions of each pixel with the method described in section~\ref{sect:data_redct}. The readout noise and the $\sigma_{\rm dark50}$ of each pixel are obtained in dark experiments described in section~\ref{subsect:dark_prpt}. Therefore, we have built a complete database of the properties of each pixel in the three areas, 29403 pixels in total. %Further studies based on these pixel-by-pixel properties are introduced below.

\begin{figure}
\begin{center}
\begin{tabular}{c}
\includegraphics[width=1.0\textwidth]{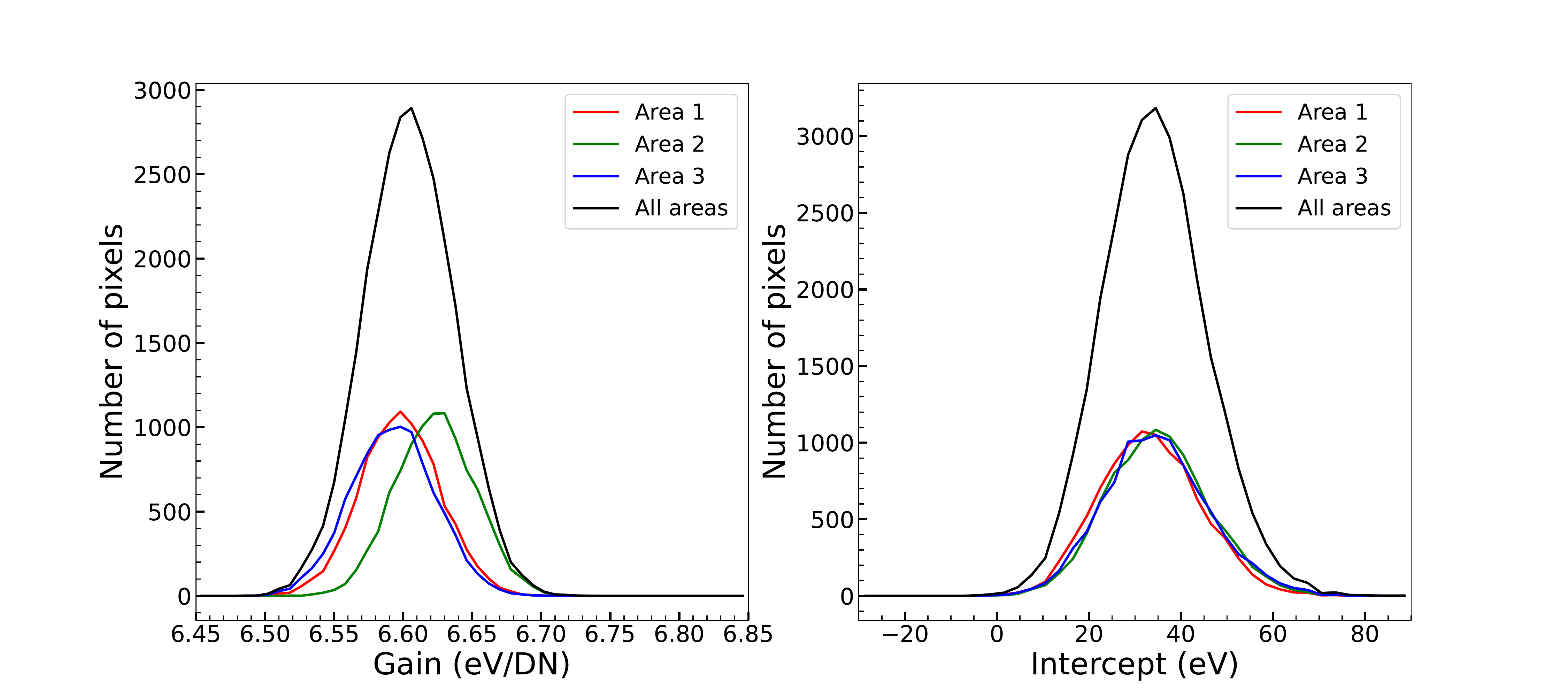}
\end{tabular}
\end{center}
\caption 
{ \label{fig:gain_dist}
The distribution of gains (left panel) and intercepts (right panel) among pixels in three areas separately (red, green, blue) and in all areas altogether (black). The gain distributions in different areas are not the same. However, the intercept distributions are similar.} The overall gain variation among all these pixels is 0.56\%. 
\end{figure}

For the gain of each pixel, the measurement error is around 0.15\%, or 0.01 eV/DN. And for the intercept measurement, the error is around 6 eV. They are sufficiently small so that we can capture the tiny variations among the gains and intercepts. The left panel of Fig.~\ref{fig:gain_dist} shows the distribution of the gains in the three areas separately and in total. In Area 1 and Area 3, the mean values of the gains are similar, around 6.59 eV/DN. However, Area 2 has a slightly higher mean value of 6.62 eV/DN. The gain variations in the three areas are not exactly the same as well: in Area 1 and Area 2 they are around 0.43\%, while in Area 3 it is around 0.68\%. This shows that there are gain differences not only among pixels in a small region, but also among different regions on the chip, which reflects the variations on small and large scales, respectively. The overall mean value comes to 6.60 eV/DN and the gain variation for pixels in all the three areas is 0.56\%. The variation of the intercepts is given by the right panel of Fig.~\ref{fig:gain_dist}. In the three areas separately and in total, the distributions of the intercepts are all similar: the means are around 34 eV and the standard deviations are around 11 eV. These non-zero intercepts are caused by the charge loss of the G0center events.

In theory, the energy resolution of a CMOS sensor is given by the following equation \citep{hull2019hybrid}:
\begin{equation}
\label{eq:fwhm}
    FWHM = 2.355\times \omega \sqrt{(\frac{\sigma_{\rm gain}\times E}{\omega})^2 + \frac{F\times E}{\omega} + \sigma_{\rm total}^2},
\end{equation}
where the pair creation energy $\omega=3.65\ \rm{eV/e^-}$ at room temperature, the Fano factor $F=0.118$ \citep{lowe2007a, rodrigues2021absolute}, $E$ is the incident photon energy, $\sigma_{\rm gain}$ is the gain variation and $\sigma_{\rm total}$ is the total noise. The dark noise at the integration time of $50\ \rm{ms}$, $\sigma_{\rm dark50}$, is included in the total noise.

\begin{figure}
\begin{center}
\begin{tabular}{c}
\includegraphics[width=0.75\textwidth]{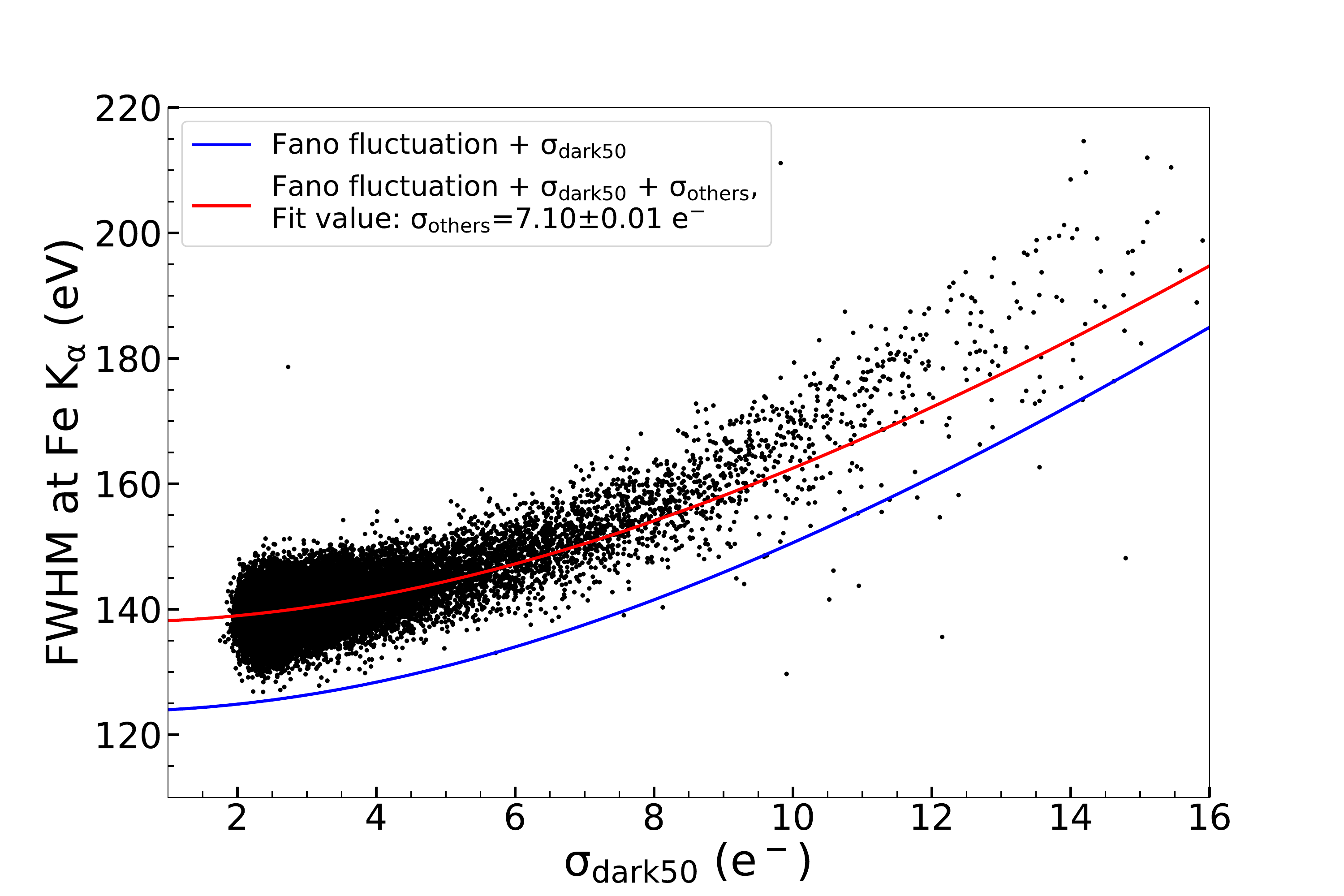}
\end{tabular}
\end{center}
\caption 
{ \label{fig:noi_res}
The relationship between the FWHM at Fe $\rm{K_\alpha}$ line and the noise at the integration time of $50\ \rm{ms}$, $\sigma_{\rm dark50}$, at the pixel level. The blue curve gives the theoretical result that is contributed by the Fano fluctuation and $\sigma_{\rm dark50}$. And the red curve gives the fitting result using Eq.~\ref{eq:fwhm_pixel}.} 
\end{figure}

At the pixel level, there is no gain variation ($\sigma_{\rm gain}=0$), and the energy resolution of a single pixel is related to the noise level. Fig.~\ref{fig:noi_res} shows this correlation between the FWHM at Fe $\rm{K_\alpha}$ and the noise at the integration time of $50\ \rm{ms}$, $\sigma_{\rm dark50}$. The blue curve gives the theoretical result that is contributed by the Fano fluctuation and $\sigma_{\rm dark50}$. The gap between the curve and the data indicates that there are other types of noise, $\sigma_{\rm others}$. Therefore, the energy resolution of a single pixel should be given by:
\begin{equation}
\label{eq:fwhm_pixel}
    FWHM = 2.355\times \omega \sqrt{\frac{F\times E}{\omega} + \sigma_{\rm dark50}^2 + \sigma_{\rm others}^2}.
\end{equation}
The data is then fitted with Eq.~\ref{eq:fwhm_pixel}, which gives $\sigma_{\rm others} = 7.1 \ \rm{e^-}$. It shows that not only the readout noise and the contribution from the dark current are important for the energy resolution of a single pixel, but some other sources of noise should be concerned.

\subsection{Gain Correction and Energy Resolution}
\begin{figure}
\begin{center}
\begin{tabular}{c}
\includegraphics[width=0.75\textwidth]{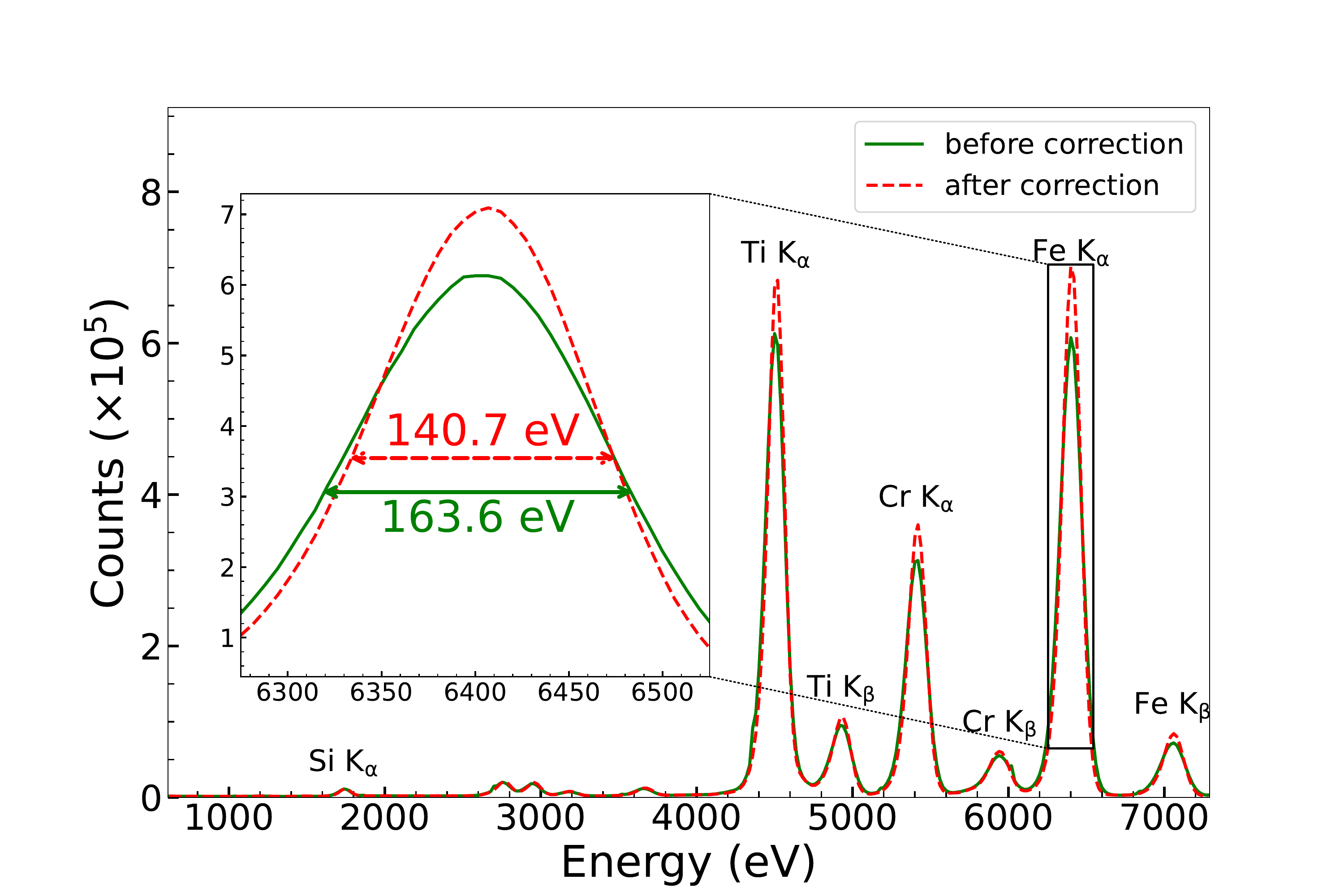}
\end{tabular}
\end{center}
\caption 
{ \label{fig:spec_compare}
The G0center spectra before (green curve) and after (red dashed curve) the gain correction. Obviously, the spectrum after the correction has higher peaks and better energy resolutions: the FWHM improves from 141.0 eV to 124.6 eV at 4.5 keV, and from 163.6 eV to 140.7 eV at 6.4 keV. These results are obtained at room temperature.} 
\end{figure}

In previous studies, we have always handled X-ray events from all pixels on a CCD or CMOS detector in the same way, using the same conversion gain across the whole chip. Based on the results of section~\ref{subsect:pixel_prop}, the gain inhomogeneity can apparently degrade the energy resolution. However, with a complete database of the pixel-by-pixel properties, we can perform this energy conversion for each of the pixels individually, which is expected to result in a more precise energy spectrum. We have done the gain correction for the G0center spectrum using the events from all three areas. According to the position of the incident photon, the conversion gain of the pixel at this position is used to convert the digital numbers into energies. The spectra before and after the correction are given in Fig.~\ref{fig:spec_compare}. Obviously, the spectrum after the correction has higher peaks and better energy resolutions: the FWHM improves from 141.0 eV to 124.6 eV at 4.5 keV, and from 163.6 eV to 140.7 eV at 6.4 keV, which are close to those of a single pixel. Note that these energy resolutions are realized at room temperature, further highlighting the advantages of the CMOS sensors.

\begin{figure}
\begin{center}
\begin{tabular}{c}
\includegraphics[width=0.75\textwidth]{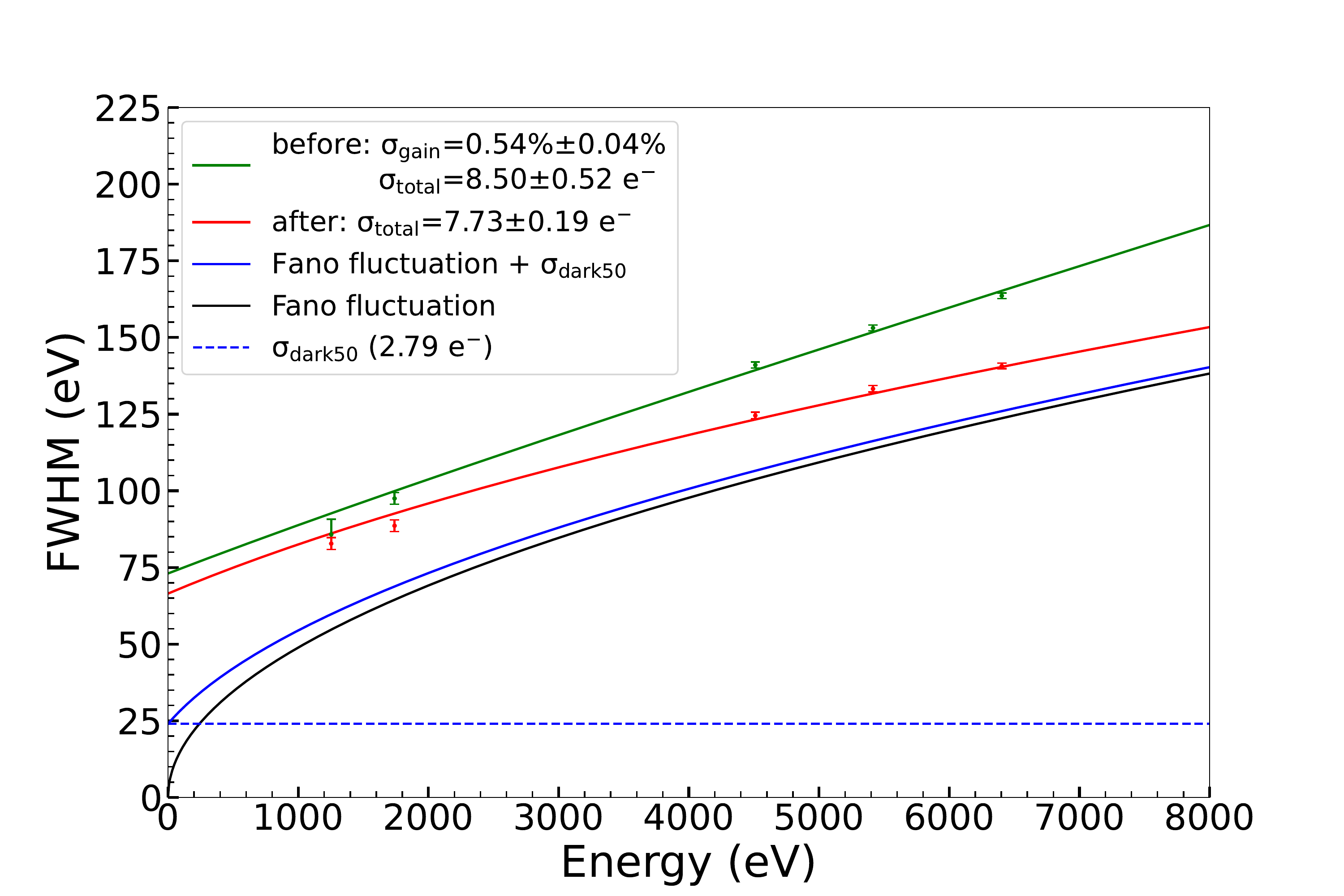}
\end{tabular}
\end{center}
\caption 
{ \label{fig:fwhm_fit}
The energy resolutions for peaks before and after the gain correction. The result of the Mg $\rm{K_\alpha}$ line at 1.25 keV is added. The energy resolutions before the correction are well fitted with Eq.~\ref{eq:fwhm} (green curve). The results after the gain correction are also well fitted with $\sigma_{\rm gain}$ set to 0 in Eq.~\ref{eq:fwhm} (red curve). The black curve gives the Fano fluctuation and the blue curve is the results with the $\sigma_{\rm dark50}$ added. The $\sigma_{\rm dark50}$ contribution is shown as the blue dashed line.} 
\end{figure}

The energy resolutions before and after the gain correction are given in Fig.\ref{fig:fwhm_fit}. The result of the Mg $\rm{K_\alpha}$ line at 1.25 keV is obtained from a Mg target exposure experiment. The energy resolutions before the correction are well fitted with Eq.~\ref{eq:fwhm} (green curve). The fitted $\sigma_{\rm gain}=0.54\%\pm0.04\%$ is consistent with the gain variation measured in this work ($0.56\%$). The energy resolutions after the gain correction are also well fitted with $\sigma_{\rm gain}$ set to 0 in Eq.~\ref{eq:fwhm} (red curve), meaning that the degradation due to the pixel inhomogeneity is basically eliminated by the gain correction. The results of Fano fluctuation and that with $\sigma_{\rm dark50}$ added are also shown. These results prove that the gain correction can significantly improve the energy resolution. The fit value of the total noise $\sigma_{\rm total}$, around 7.7 $\rm{e^-}$, is much higher than $\sigma_{\rm dark50}$, which is only 2.8 $\rm{e^-}$, indicating a $\sigma_{\rm others}$ of around 7.2 $\rm{e^-}$. This is consistent with the fit value of 7.1 $\rm{e^-}$ in Fig.~\ref{fig:noi_res}. $\sigma_{\rm others}$ may come from varies of sources. The image lag, which is caused by the residual charges left by the previous frame, can influence the exposure of the next frame and can be one possible noise source. Noise may also come from the charge loss caused by the recombination in long-distance diffusion and the absorption by edges of pixels. We will study these noises further in the future.

\section{Conclusions}
\label{sect:concls}
CMOS detectors have great potential in X-ray astronomy due to their excellent performances. Nowadays, the readout noise of a typical scientific CMOS sensor can be less than 5 $\rm{e^-}$ and some even reach 1 $\rm{e^-}$. However, the typical energy resolution of sCMOS sensors is worse than the Fano fluctuation limit even when the readout noise is taken into account. The inhomogeneity between pixels may be one of the factors. To study this problem and improve the energy resolution, we investigated the properties of the sCMOS sensor at the level of individual pixels. 

Using a customized sCMOS sensor GSENSE\-1516\-BSI, we established a complete database of pixel-by-pixel properties of around 30 thousand pixels in the areas we selected on the chip. These properties include the noise level, the conversion gain and the energy resolutions of each pixel. At room temperature, the energy resolution of a single pixel can reach 140 eV at 6.4 keV. The positive correlation between the noise and the energy resolution supports the notion that the noise, including the readout noise and the contribution from the dark current, is an important factor for the energy resolution of a single pixel. However, comparison with theoretical results indicates that other noise sources must exist to explain the energy resolution. These noises will be studied further in the future. The gain variation, which exists not only in one area but also among areas, is measured to be 0.56\% for all pixels, consistent with the theoretical fit result. In the future, this variation is expected to be reduced by advances in manufacturing techniques.

Based on the properties of each pixel, we can do a thorough pixel-by-pixel gain correction over the whole chip to eliminate the degradation of energy resolution caused by the gain variation. We used three regions, each consisting of $99\times99$ pixels, to demonstrate this method and obtained energy resolutions of 124.6 eV at 4.5 keV and 140.7 eV at 6.4 keV, respectively. In principle, this gain correction can be applied to the whole chip, which can make the energy resolution of the chip close to the single-pixel energy resolution of around 140 eV at 6.4 keV. In future work, we will apply this method to events of all grades, instead of 1-pixel events only. The energy resolutions above are realized at room temperature, which demonstrates the advantages of CMOS sensors. For future spectroscopic applications in X-ray, CMOS sensors are an excellent choice, and a thorough pixel-by-pixel gain correction is recommended to achieve a better performance.

\begin{acknowledgments}
The authors thank the referee for his/her helpful comments. This work is supported by the National Natural Science Foundation of China (grant No. 12173055) and the Chinese Academy of Sciences (grant Nos. XDA15310100, XDA15310300, XDA15052100).
\end{acknowledgments}






\bibliography{reference.bib}{}
\bibliographystyle{aasjournal}



\end{document}